\documentclass[final,smallcondensed,numbook,dvipsnames]{svjour3}

\usepackage[english]{babel}							
\usepackage[utf8]{inputenc}
\usepackage{amssymb}
\usepackage{graphicx}
\usepackage{amsmath}
\usepackage{multirow} 
\usepackage[parfill]{parskip} 
\usepackage{cancel,mathtools,empheq,latexsym}
\usepackage{subfig,epsfig,tikz,float}		        
\usepackage{booktabs,multicol,multirow,tabularx,array} 
\usepackage{bm}
\usepackage{mathrsfs}
\usepackage{comment} 
\usepackage{changepage} 
\usepackage[font={it}]{caption} 
\usepackage{hyperref}
\hypersetup{
    colorlinks=true,       
    linkcolor=red,          
    citecolor=blue,        
    filecolor=magenta,      
    urlcolor=cyan           
}

\title{Lambert's problem in orbital dynamics: a self--contained introduction}

\author{Lenox Helene Baloglou
    \thanks{All authors were partially funded by the United States National Science Foundation through the grant NSF-DMS-2137305. Lenox H. Baloglou is thankful for the generous support of the University of Arizona's RII-Sponsored Campuswide Undergraduate Student--Initiated Original Research Program.} \and
        Parneet Gill
    \and
Tonatiuh S\'anchez-Vizuet
    }
\institute{Lenox Helene Baloglou
            \at Department of Mathematics, The University of Arizona, USA.
            \email{lenoxbaloglou@arizona.edu} \and
           Parneet Gill
            \at Department of Mathematics, The University of Arizona, USA.
            \email{pgill@arizona.edu} \and
           Tonatiuh S\'anchez-Vizuet
            \at Department of Mathematics, The University of Arizona, USA. ORCID: https://orcid.org/0000-0002-8930-2798.
            \email{tonatiuh@arizona.edu}
    }

\authorrunning{Baloglou, Gill, S\'anchez-Vizuet}

\begin{document}

\maketitle

\begin{abstract}
Lambert's problem is a classical boundary value problem in analytical mechanics. It arises when trying to determine the energy required to place a particle, subject to a central gravitational potential, in a \textit{free fall} trajectory connecting two given points on a desired travel time. Due to its mathematical beauty and its relevance in aerospace engineering, it has been and remains the object of attention of countless engineers, mathematicians (pure and applied), and physicists seeking to produce efficient solution algorithms. In this expository article, didactic in nature, we present a unified and comprehensive derivation that assumes only a minimal background in physics and mathematics. We focus on the simplest unperturbed case and carefully develop the argument for elliptical trajectories. The goal is to provide a single reference that can serve as an accelerated introduction for students and researchers interested in a quick introduction to the subject.
   
\keywords{Classical mechanics \and Orbital transfer \and Orbital dynamics \and Keplerian dynamics \and Motion under a central potential \and Lambert problem.}
\subclass{70-01 \and 85-01 \and 70M20 \and 70F16 \and 70F05.}
\end{abstract}

\section{Introduction}

The goal of this expository article is didactic: we aim to provide a unified reference for readers who are comfortable with calculus, geometry and differential equations---but may not necessarily have strong background in physics---and are interested in the field of orbital dynamics in general and the Lambert problem in particular. All of the topics in this text are available in the literature---and have been available for a long while---however, a novice reader often has to spend a significant amount of time and effort jumping from one reference to another just to piece together a reasonably strong background before being able to tackle research--level questions. This was in fact the experience the two first authors---graduate students in applied mathematics interested in the use of surrogate methods for the solution of Lambert's problem \cite{Baloglou2025,BaGiSa2025}; this text grew out of their background--building efforts. Given the instructional goal and our target readership, we provide significantly more details in the calculations than what a typical research article---and even some textbooks---would do. We believe that this approach (and this reference itself) will be especially useful to students---and even researchers---in physics, astronomy, aerospace engineering and applied mathematics who find themselves in need of an accelerated introduction to the subject. 

Lambert's problem remains a very active field of research both from the purely mathematical point of view and from the practical application--oriented one. We hope that, with this accelerated minimal introduction, we can save the reader some time and provide them with the foundations that they will need to take a deeper dive into the multiple theoretical and practical complications of the field. Some of these include dealing with drag \cite{Urena2023}, multiple revolutions and gravitational perturbations \cite{Armellin2018,arora2013,Eagle,Panicucci2018,Woollands2017}, uncertainty and stochasticity \cite{Adurthi2020,Schumacher2015,Teter2025,Zhang2018}, and the development of robust and efficient computational algorithms \cite{BaGiSa2025,Gueho2020,Thompson2020,Yang2022}, among many others.
 
Part of the beauty of orbital dynamics is that it constitutes a natural connection between geometry, physics and calculus and our exposition attempts to emphasize this connection. In Section \ref{sec:2}, we start by providing a brief introduction to the Apollonian description of conic sections and their analytic geometry. This description is not often covered in elementary classes and yet it provides the most natural framework for Keplerian dynamics. We then move on to a minimal introduction to the physical concepts required. This material is part of the standard curriculum of an introductory physics class, but may not be familiar to a reader with a background in mathematics or computer science. Section \ref{sec:3} pertains to the more advanced subject of Keplerian dynamics. This subject is covered in classes of advanced mechanics (typically going by the names of theoretical or analytical mechanics) and once again we strive to synthesize just the necessary concepts to spare the reader the need for a semester--long incursion into this beautiful field. Having collected all the necessary material, Section \ref{sec:Lambert} finally introduces Lambert's problem and provides a careful derivation of Lagrange's solution to it, all of which is typically found---although with a much terser exposition---in specialized texts and articles on orbital dynamics.  

\textbf{A few words about notation.\\}
Throughout the text, we will use the following notational conventions:
\begin{itemize}
\item Scalar--valued quantities will be denoted with lightface and vector--valued quantities with boldface. Hence, $r$ is a scalar and $\boldsymbol r$ is a vector.

\item Angular variables will be denoted by Greek letters.

\item If $P$ and $Q$ are two points in space, the length of the straight line segment connecting them will be denoted by $|PQ|$.

\item If $P$ is a point and $L$ is a line, then $|PL|$ should be understood as the length of the line segment going through the point $P$ and intersecting the line $L$ at a right angle---this is the shortest distance between $P$ and $L$.

\item If $\boldsymbol L$ is a vector, then $|\boldsymbol L|$ will denote its Euclidean magnitude.

\item We will place a ``hat" \,  $\widehat{}$\, on top of a vector whenever the Euclidean magnitude of said vector is equal to one. Therefore $\widehat{\boldsymbol r}$ implies that $|\boldsymbol r| = 1$.

\item The symbol $:=$ denotes the fact that the quantity on the left is \textit{defined} to be equal to the quantity on the right.

\item Differentiation with respect to time will often (but not always) be denoted by a dot on top of a variable; the number of dots denotes the order of the derivative, i.e. ${\displaystyle \dot x : = \tfrac{d}{dt}x}$ and ${\displaystyle \ddot x : = \tfrac{d^2}{dt^2} x}$.
\end{itemize}
\section{Preliminary background}\label{sec:2}

\subsection{\textbf{Conic sections}}

This section is a refresher---or crash course---on analytic geometry for conical sections. Since the natural coordinate system for the description of a particle moving under a central potential is a focus--centered polar system, except for a few comments, we will not delve on the Cartesian description. We start by introducing some simple notation. For two points $P_1$ and $P_2$ in the Euclidean plane we will denote by $|P_1P_2|$ the length of the straight segment connecting them (i.e. their distance), while if $L$ is a line and $P$ a point, the length of the shortest line segment passing through $P$ and intersecting $L$ will be denoted by $|PL|$. 

\textbf{Apollonian conics.}
There are several equivalent definitions of the conic sections. Perhaps the most well--known are the association between them and planar intersections of cones, and the characterization in terms of quadratic forms. However, the most useful one for our purposes will be the definition due to the Greek geometer Apollonius of Perga \cite{Apollonius}. The reason why we will use this definition is that it encompasses almost all of the definitions for conics in one single condition, in a way that is equivalent to the \textit{standard} definition in terms of quadratic forms \cite[\,Theorem 2.1]{GlStOd2016}.

\begin{figure}
\centering
\begin{tabular}{cc}
 & \includegraphics[height = 0.225\linewidth]{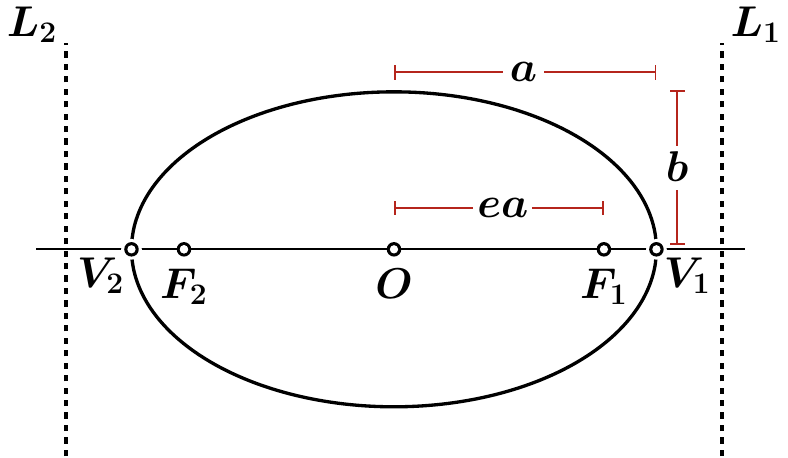}\\[4ex]
\multirow[t]{2}{*}{\includegraphics[height = 0.44\linewidth]{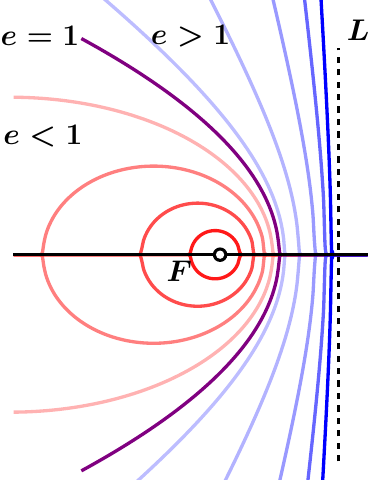}} \hspace{1.5cm} & \includegraphics[height = 0.2\linewidth]{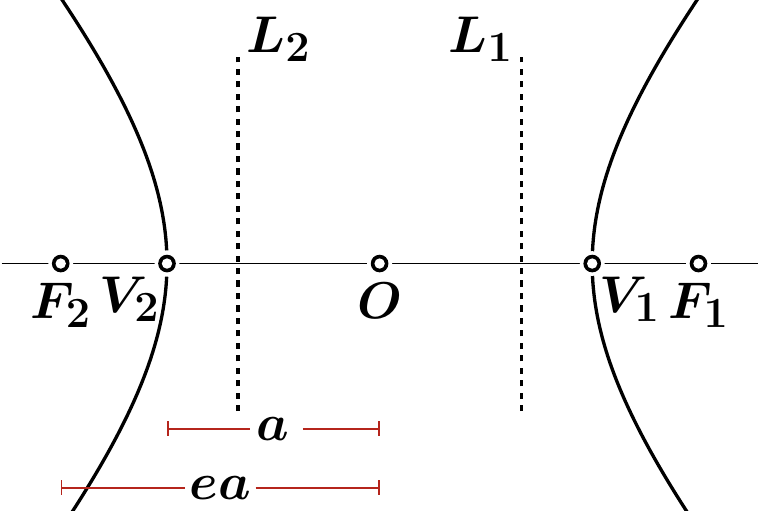}
\end{tabular}
\caption{Left: Conical sections for different values of the eccentricity $e$. Shades of red represent values of $0<e<1$ resulting in ellipses; the color fades towards one. Shades of blue represent values of $e>1$ resulting in hyperbolae with the color fading towards one. The parabola, plotted in violet, is the limiting case $e=1$. The directrix $L$ is plotted as a dashed line, the focus $F$ is marked by an open circle. Right: Relevant geometric markers for ellipses and hyperbolae.}\label{fig:ApollonianConics}
\end{figure}

\begin{definition}[\cite{besant1890,GlStOd2016}]\label{def:Conic}
Consider a point $F$ in the Euclidean plane that we will call \textit{focus}, and a straight line $L$ not passing through $F$ that we will call \textit{directrix}. For any $e>0$, the set of points
\[
\mathcal C :=\{P \in\mathbb R^2: |PF|=e|PL|\}
\]
is called a \textit{conic} with associated focus $F$ and directrix $L$. Moreover, we say that the conic $\mathcal C$ is: an ellipse, if $0<e<1$, a parabola if $e=1$, or a hyperbola if $e>1$.
\end{definition}
The line connecting $F$ with its directrix at a right angle (depicted in solid black in Figure \ref{fig:ApollonianConics}) defines an axis of symmetry of the conic, known as the \textit{major} or \textit{principal} axis. The point $V$ along the segment $FL$ and belonging to the conic is known as the \textit{vertex}. As we shall now see, ellipses and hyperbolas have a second axis of symmetry and therefore a second focus and directrix, however the eccentricity (with respect to the second pair focus/directrix) remains unchanged. The following argument has been adapted from \cite[\,Lemma 2.1.1]{GlStOd2016}.

Let's consider a directrix/focus pair where the directrix $L$ is oriented in the vertical direction and the focus $F$ is located to its left at some distance $|FL|$. Placing a Cartesian frame of reference with $x$ axis parallel to the major axis of symmetry and with its $y$ axis parallel to the directrix $L$, we see that the focus has coordinates $(-|FL|,0)$. Therefore, for a point with coordinates $(x,y)$, the condition in Definition \eqref{def:Conic} can be expressed as
\begin{equation}\label{eq:ApollonianDefinition}
(x+|FL|)^2 + y^2 = e^2x^2.
\end{equation}
If $e=1$ (the parabolic case) the quadratic term on $x$ drops from the expression above, and it becomes clear that there will be only one point in the conic along the symmetry axis (i.e. with $y=0$). On the other hand, assuming that $e\neq 1$ the quadratic formula yields
\begin{equation}\label{eq:Xcoord}
x  = \frac{-|FL|\pm\sqrt{|FL|^2-(1-e^2)(y^2 + |FL|^2)}}{1-e^2}.
\end{equation}
Letting $y=0$ above, we conclude that there are two distinct points along the axis of symmetry that belong to the conic. In other words, the conic will have two vertices $V_1$ and $V_2$. The length of the line segment connecting them is given by
\[
|V_1V_2| = \frac{2\,e|FL|}{\left|1-e^2\right|}.
\]
The \textit{semi--major axis}, familiar from the Cartesian description of conics, is defined to be half of this length, namely
\begin{equation}\label{eq:SemiMajor}
a: = \frac{e\,|FL|}{\left|1-e^2\right|}.
\end{equation}
For all values of $y$ for which Equation \eqref{eq:Xcoord} is well defined, we see that there will be two points in the conic with the same $y$ coordinate. Since the duplicity comes from the two branches of the square root, we conclude that there is second axis of symmetry---that we shall denote the \textit{minor} or \textit{secondary} axis---located along the vertical line
\[
x_O = -\frac{|FL|}{1-e^2}.
\]
The point where the axes of symmetry intersect each other is known as the \textit{center} of the conic and we shall denote it by
\[
O := \left(-\frac{|FL|}{1-e^2},\,0\right).
\]
Setting $x=x_O$ in \eqref{eq:Xcoord} it is possible to solve for $y$ to obtain
\[
y = \pm|FL|\sqrt{\frac{e^2}{1-e^2}},
\]
which, in the elliptic case $0<e<1$, yields the value of the $y$ coordinate of the point in the ellipse located vertically above/below the center $O$. The length of the \textit{semi--minor} axis is defined as the distance
\begin{equation}\label{eq:SemiMinor}
b := |FL|\sqrt{\frac{e^2}{1-e^2}}.
\end{equation}
From the discussion above, it follows that the distance between the center $O$ and either of the foci is given by
\begin{equation}\label{eq:OF}
|OF| = \left|\,|FL|- \frac{|FL|}{1-e^2}\right| = \frac{e^2|FL|}{\left|1-e^2\right|} \underbrace{=}_{\text{By }\eqref{eq:SemiMajor}} e a.
\end{equation}
From the final equality above, we can also write the distance between the focus and the directrix in terms of the length of the semi--major axis and the eccentricity as
\[
|FL| = \frac{a}{e}|1-e^2|.
\]
Finally, solving for $|FL|$ in \eqref{eq:SemiMajor} and substituting the result in \eqref{eq:SemiMinor} yields
\begin{equation}\label{eq:AandB}
b = \frac{a|1-e^2|}{e}\,\sqrt{\frac{e^2}{1-e^2}} = a\sqrt{1-e^2}.
\end{equation}
%
\textbf{The circle.}
The reader must have noticed that the circle is missing from the definition above. We will now see that it can be understood as a special case of an ellipse with eccentricity $e=0$. However, if we simply let $e=0$ in Definition \ref{def:Conic}, we would only recover the degenerate case of a single point: the focus. Any attempt at using Equation \eqref{eq:ApollonianDefinition} to obtain the circle as a limiting case of the Apollonian definition will face the same shortcoming. Hence, to include the circle as a limiting case, we will have to make the additional assumption that the length of the semi--major axis, $a$, is a fixed predetermined parameter independent from the eccentricity.

We will first show that the Apollonian definition implies the well--known characterization of ellipses in terms of the constant sum of distances from the foci. With that in mind, we go back to the elliptic case and observe that the existence of the second (vertical) axis of symmetry at $x_O$ implies:
\begin{enumerate}
\item The existence of a second focus located along the major axis of symmetry with horizontal coordinate given by $|LF_2| = x_O + ae$ so that the two foci are located at the points
\[
F_1 = (x_O-ea,0) \qquad \text{ and } \qquad F_2 = (x_O+ea,0).
\]
\item That, for every point on the ellipse, there is a second point in the ellipse sharing the same $y$ coordinate and symmetrically located horizontally with respect to $x_O$. Namely, if
\[
P_1 = (x,y)\in\mathcal C  \qquad \text{ then } \qquad P_2 = \left(x + 2(x_O - x),y\right) = (2x_O - x,y)\in\mathcal C.
\]
\end{enumerate}
From the two points above it follows that
\[
|P_1F_2| = |x - (x_O + ea)| = |x_O+ea-x| = |(2x_O-x)-(x_O-ea)| = |P_2F_1|,
\]
and therefore
\[
|P_1F_2| + |P_1F_1| = \underbrace{|P_2F_1| + |P_1F_1| = e(|P_2L| + |P_1L|)}_{ \text{\scriptsize From the definition }\ref{def:Conic}} = e\left((2x_O - x) + x \right) = \frac{2e|FL|}{1-e^2} = 2a.
\]
Summarizing, if a point $P$ belongs to the ellipse $\mathcal C$, it satisfies the constant sum of distances property
\begin{equation}\label{eq:ConstantSum}
|P_1F_2| + |P_1F_1| = 2a.
\end{equation}
On the other hand, the distance between the foci is given by $|F_1F_2| = 2ae$, while the distance between the foci and the center is $|F_1O|=|F_2O|=ae$. Therefore, if $e=0$ it follows that $F_1=F_2=O$, and therefore the constant sum of distances property implies that
\[
|PO| = a.
\]
This is, of course, the definition of a circle with radius $a$, which shows that the circle can be considered as the limiting case of ellipses with vanishing eccentricity.

\textbf{Focal equation of a conic.}
Placing a reference frame at the focus $F$ with the horizontal axis parallel to the line segment $FL$, it is possible to describe the position of a point $P$ in terms of its distance, $r$, from the focus and the angle, $\theta$, that the vector connecting $F$ to $P$ makes with the horizontal axis. For historical reasons, within the context of celestial mechanics, the angle $\theta$ is known as the \textit{true anomaly}.

\begin{figure}
\centering
\includegraphics[height = 0.25\linewidth]{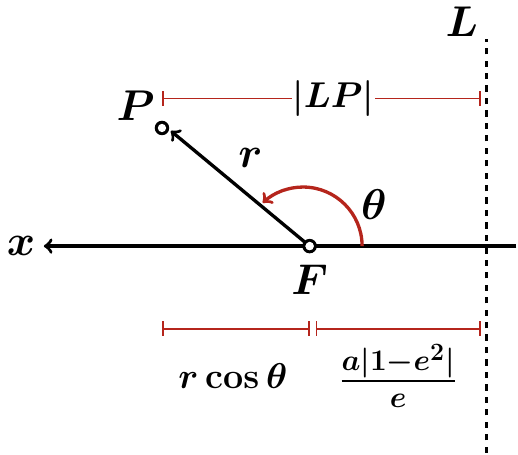}
\caption{Polar description of a point $P$ in terms of the semi--major axis $a$ and eccentricity $e$.}\label{fig:PolarEquationGeometry}
\end{figure}

As depicted in Figure \ref{fig:PolarEquationGeometry}, using this focus--centered polar coordinate frame, the condition from Definition \ref{def:Conic} can be expressed as
\[
r = e|LP| = e\left(|FL| - r\cos\theta\right).
\]
If we solve for $r$ in the expression above, we obtain
\begin{subequations}\label{eq:PolarConic}
\begin{equation}\label{eq:PolarConicA}
r = \frac{e|FL|}{1+e\cos\theta}
\end{equation}
which expresses the relationship between the radial distance $r$ and the polar angle $\theta$ in terms of the Apollonian parameters $|FL|$ and $e$ (recall that the Apollonian definition of conics uses only the directrix $L$, the focus $F$ and the eccentricity). The expression above, known as the \textit{focal equation of a conic} \cite{Arnold1978}, does not include the circle: letting $e=0$ above collapses the conic into a point at the focus. This should not come as a surprise, since the Apollonian definition used to derive this expression does not include the circle.

However, just as we did before, if we assume that the semi--major axis $a$ is an additional parameter independent of $e$, we can use \eqref{eq:SemiMajor} to eliminate $|FL|$ from the expression above and obtain
\begin{equation}\label{eq:PolarConicB}
r = \frac{a|1-e^2|}{1+e\cos\theta}.
\end{equation}
\end{subequations}
In this formulation, letting $e=0$ results in the equation of a circle of radius $a$. However, the expression above does not include the parabolic case, as for $e=1$ the expression collapses into a point. This was to be expected, as the semi--major axis is not defined for parabolas. Since we will focus on the elliptic and circular cases later on, we will prefer this last expression for the conics.

\begin{remark}[On the sign of $e$ in the focal equation]\label{eq:remarkSignE}
 Our derivation of equations \eqref{eq:PolarConic} used the geometric construction from Figure \ref{fig:PolarEquationGeometry}, where the origin of the coordinate system is located at the \textit{right} focus of the ellipse: the one labeled as $F_1$ in the top right panel of Figure \ref{fig:ApollonianConics}, and the directrix is the line $x=|FL|$. However, it is possible and equally valid to place the frame of reference at the \textit{left} focus (labeled $F_2$ in Figure \ref{fig:ApollonianConics}) so that the directrix is the line $x=-|FL|$. In that case, if the angle is still measured from the positive horizontal axis, the equations derived by the same procedure will be analogous to \eqref{eq:PolarConic} with the only difference being that the eccentricity in the denominator will appear with the opposite sign. Hence, depending on the references, the denominator of \eqref{eq:PolarConic} can be equal to $1+e\cos\theta$, as in our case, or to $1-e\cos\theta$. Some references go as far as writing the denominator as $1\pm e\cos\theta$. The reader should be aware of this possible discrepancy and how to resolve it. 
\end{remark}

\subsection{Elementary concepts from mechanics}
The following elements from theoretical mechanics can be found in any standard reference, such as \cite{Morin2012,Scheck2010,ThMa2014}. For the benefit of the reader with no previous familiarity with physics, we collect here just the concepts that will be absolutely essential for our exposition.

\textbf{Governing laws.}
We are ultimately interested in describing the motion of a point particle (i.e. one that can be idealized as occupying no volume) with mass $m$ under the action of a gravitational force $\boldsymbol F$ produced by an attracting body of mass $M$. We will refer to the body with mass $M$ as the \textit{attractive center} and will also consider it to be a point. We will place a reference frame at the location of the attracting center and will determine the position of the particle of mass $m$ by the vector $\boldsymbol r$, anchored at the origin, connecting the two particles. We will denote the magnitude of the position vector by $r:=|\boldsymbol r|$. As is customary in mechanics and dynamical systems, we will use a dot to represent time differentiation, and will denote the velocity of the particle by $\dot{\boldsymbol r}$. Under these circumstances, the mathematical description of the motion of the particle will be determined by
\begin{alignat*}{6}
m\ddot{\boldsymbol r} =\,& \boldsymbol F && \text{(Newton's second law of motion)},\\
\boldsymbol F =\,& -\frac{GMm}{r^3} \boldsymbol r \qquad\qquad&& \text{(Newton's universal law of gravitation)}.
\end{alignat*}
Above, the quantity $G$ is a numerical constant known as the \textit{gravitational constant} and the minus sign appearing in the second equation indicates that the gravitational force is attractive. Given that $\boldsymbol F$ depends only on the position $\boldsymbol r$ of the particle relative to the attractive center, it is referred to as a \textit{central force}. Moreover, the reader will find it easy to verify that the force can be expressed as the gradient of a scalar function in the form:
\begin{equation}\label{eq:PotentialEnergy}
\boldsymbol F(\boldsymbol r) = -\nabla U(r), \qquad \text{ where } \qquad U(r) := -\frac{GMm}{r}.
\end{equation}
The scalar function $U$ is referred to as \textit{potential energy}, and forces that can be expressed as the gradient of a scalar quantity are said to be \textit{conservative}. From the two laws of motion above, we see that a particle under the influence of a gravitational force will be then constrained to satisfy the second order differential equation
\begin{equation}\label{eq:SecondOrderEquation}
\ddot{\boldsymbol{r}} = -\frac{GM}{r^3}\boldsymbol r.
\end{equation}  
\textit{A priori}, the expression above is a system of three second order differential equations (one for each component of the position vector $\boldsymbol r$) that, given the appropriate initial conditions, completely determines the movement of a particle under the influence of the attractive center. Rather than setting out to integrate these equations, in what follows we will exploit the physical properties of the system to reduce it first into a two dimensional second order vector problem, and then into a first order scalar problem. 

\textbf{Conservation of angular momentum.}
For a point particle with mass $m$, the vector quantity
\[
\boldsymbol L : = \boldsymbol r \times m \boldsymbol \dot{\boldsymbol r}\,
\]
is known as the \textit{angular momentum}. The symbol ``$\times$" above denotes the standard vector or cross product satisfying for all $\boldsymbol u,\boldsymbol v\in\mathbb R^3$:
\begin{equation}\label{eq:CrossProd}
\boldsymbol u\times\boldsymbol v = |\boldsymbol u||\boldsymbol v|\sin\phi\,\boldsymbol n,
\end{equation}
where $\phi$ is the angle between $\boldsymbol u$ and $\boldsymbol v$, and $\boldsymbol n$ is the unitary vector normal to the plane determined by $\boldsymbol u$ and $\boldsymbol v$ in the positive direction.

From its definition, the angular momentum is perpendicular to both the particle's position and velocity and, at any given time it determines the instantaneous plane of motion of a particle. We will now compute the rate of change of the angular momentum for a particle under the action of a gravitational force:
\[
\dot{\boldsymbol L} = \frac{d}{dt}\left(\boldsymbol r \times m\dot{\boldsymbol r}\right) = \dot{\boldsymbol r}\times m\dot{\boldsymbol r} + \boldsymbol r \times m \ddot{\boldsymbol r} \underbrace{=}_{\text{\scriptsize By \eqref{eq:SecondOrderEquation}}} \dot{\boldsymbol r} \times m \dot{\boldsymbol r} + \boldsymbol r \times \left(-\frac{GMm}{r^3}\boldsymbol r\right) = \boldsymbol 0,
\]
where we used the fact that the cross product vanishes when its arguments are multiples of each other. The result above is known as the \textit{conservation of angular momentum} and has long reaching consequences. The first one is that, due to $\boldsymbol L$ being normal to the plane of motion,
\begin{adjustwidth}{5cm}{0cm}\textit{
under the action of a gravitational potential, the motion of a particle is constrained to the fixed plane determined by its initial position and velocity.
}
\end{adjustwidth}
This fact has the effect of reducing the dimensionality of the problem from three to two dimensions, allowing us to use a polar coordinate system.

\textbf{Position, velocity, and acceleration in polar coordinates}
In view of the first geometric consequence of the conservation of angular momentum, we will assume that the plane of motion coincides with the Euclidean $xy$ plane and will express the position vector $\boldsymbol r$ in polar coordinates in terms of its magnitude $r:=|\boldsymbol r|$ and the angle $\theta$ that it makes with respect to the positive $x$--axis.

\begin{figure}[tb]
\centering
\includegraphics[height = 0.2\linewidth]{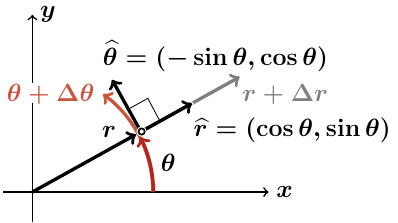} \qquad\qquad\qquad
\includegraphics[height = 0.2\linewidth]{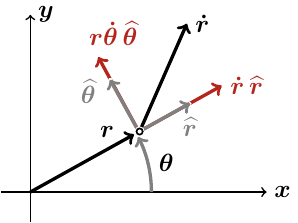}
\caption{Left: For a fixed angle $\theta$ the polar basis vectors are orthogonal and point in the direction of growth $(1+\Delta)\boldsymbol r$ (in gray) and $\theta+\Delta\theta$ (in red). Right: The velocity of a particle, $\dot{\boldsymbol r}$, can be decomposed in radial and angular components.}\label{fig:UnitVectors}
\end{figure}

The polar basis vectors $\widehat{\boldsymbol r}$ and $\widehat{\boldsymbol \theta}$ are defined as the unit vectors pointing in the directions of purely radial, $(1+\Delta r)\boldsymbol{r}$,  and purely angular, $\theta+\Delta\theta$, increment respectively. We can obtain an explicit expression for $\widehat{\boldsymbol r}$ by expressing $\boldsymbol r = (r\cos\theta,r\sin\theta)$ and inferring from the definition that
\[
 \widehat{\boldsymbol r} = \frac{(1+\Delta r)\boldsymbol{r}}{|(1+\Delta r)\boldsymbol{r}|} = \frac{\left(r\cos\theta,r\sin\theta\right)}{r} = \left(\cos\theta,\sin\theta\right).
\]
From the expression above we can obtain the one for $\widehat{\boldsymbol\theta}$ by noting that, as depicted in the left panel of Figure \ref{fig:UnitVectors}, the direction of angular growth is perpendicular to the unit radial vector, and therefore
\[
\widehat{\boldsymbol\theta} = (-\sin\theta,\cos\theta).
\]
A simple computation shows that
\begin{equation}\label{eq:DerivativeBasis}
\frac{d}{dt}\widehat{\boldsymbol r} = \dot\theta\widehat{\boldsymbol\theta} \qquad \text{ and } \qquad \frac{d}{dt}\widehat{\boldsymbol \theta} = -\dot\theta\widehat{\boldsymbol r}.
\end{equation}
Unlike their Cartesian counterparts, these unit vectors are not constant, as an angle $\theta$ is needed in order to determine the basis. Hence, fixing the polar coordinate system at the angle $\theta$, and using the identities in \eqref{eq:DerivativeBasis}, we can write the position, velocity and acceleration of a particle with Cartesian coordinates $(r\cos\theta,r\sin\theta) $ as
\begin{subequations}\label{eq:polarquantities}
\begin{align}
\label{eq:r}
\boldsymbol r =\,& r\,\widehat{\boldsymbol r}, \\
\label{eq:rdot}
\dot{\boldsymbol r} =\,& \dot r\,\widehat{\boldsymbol r} + r\,\dot\theta\,\widehat{\boldsymbol\theta},\\
\label{eq:rddot}
\ddot{\boldsymbol r} =\,& (\ddot r - r\dot\theta^2)\widehat{\boldsymbol r} + (2\dot r\dot\theta + r\ddot\theta)\widehat{\boldsymbol\theta}.
\end{align}
\end{subequations}

The quantities $\dot r$ and $\dot \theta$ are known respectively as the radial and angular velocities and are depicted in the right panel of Figure \ref{fig:UnitVectors}.

%
\textbf{Conservation of energy.}
The kinetic energy of a particle with mass $m$ moving with velocity $\dot{\boldsymbol r}$ is defined as
\[
T: = \frac{1}{2}m|\dot{\boldsymbol r}|^2 = \frac{1}{2}m(\dot r^2 + r^2\dot\theta^2)\, 
\]
where we used Equation \eqref{eq:rdot} and the fact that the vectors  $\widehat{\boldsymbol r}$ and $\widehat{\boldsymbol\theta}$ are perpendicular. The first term above is due to the radial component of the velocity---and is therefore known as the radial kinetic energy---while the second one is due to the angular component---and is referred to as the rotational energy. 

The sum of the kinetic energy and the gravitational potential energy introduced in equation \eqref{eq:PotentialEnergy}, constitutes the total energy of the system. We will denote it as
\[
E:= T + U =   \frac{1}{2}m(\dot r^2 + r^2\dot\theta^2) -\frac{GMm}{r}.
\]
Under a gravitational potential we have  
\begin{alignat}{6}
\nonumber
\frac{d}{dt} E =\,& \frac{d}{dt}\left(  \frac{1}{2}m(\dot r^2 + r^2\dot\theta^2) -\frac{GMm}{r}\right) &&\\
\nonumber
=\,&   \frac{1}{2}m(2\dot r\ddot r + 2r\dot r\dot\theta^2 + 2r^2\dot\theta\ddot\theta) +\frac{GMm}{r^2}\dot r &&\\
\label{eq:Energy1}
=\,&   m(\dot r\ddot r + r\dot r\dot\theta^2 + r^2\dot\theta\ddot\theta) +\frac{GMm}{r^2}\dot r.
\end{alignat}
Using the polar expressions of the position, velocity, and acceleration \eqref{eq:polarquantities} and the fact that $\widehat{\boldsymbol r}\cdot\widehat{\boldsymbol \theta} = 0$, it is easy to verify that
\[
m\dot{\boldsymbol r}\cdot \ddot{\boldsymbol r} = m(\dot r\ddot r + r\dot r\dot\theta^2 + r^2\dot\theta\ddot\theta),
\]
while from the second order equation of motion \eqref{eq:SecondOrderEquation} and \eqref{eq:polarquantities} we see that
\[
m\dot{\boldsymbol r}\cdot \ddot{\boldsymbol r} = -\frac{GMm}{r^3}\boldsymbol r\cdot\dot{\boldsymbol r} = -\frac{GMm}{r^2} \dot{r}.
\]
Substituting the last two results into \eqref{eq:Energy1} proves that, under a gravitational potential, the energy $E$ is constant over time. 

\textit{Remark.} The standard way of proving conservation of energy is to start from the kinetic term and compute
\[
\frac{d}{dt}\left(\frac{1}{2}m|\dot{\boldsymbol r}|^2\right) = m\dot{\boldsymbol r}\cdot\ddot{\boldsymbol r} \underbrace{=}_{\text{\scriptsize By \eqref{eq:SecondOrderEquation}}} -\frac{GMm}{|\boldsymbol r|^3}\boldsymbol r\cdot\dot{\boldsymbol r} = \frac{d}{dt}\left(\frac{GMm}{|\boldsymbol r|}\right),
\]
from which it follows that
\[
\frac{d}{dt}\left(\frac{1}{2}m|\dot{\boldsymbol  r}|^2 -\frac{GMm}{|\boldsymbol r|}\right) = \frac{d}{dt}\left(T + U \right) = \frac{d}{dt}E = 0.
\]

\section{Motion under a gravitational potential}\label{sec:3}

\subsection{Equations of motion}

As we mentioned before, the motion of a particle under a gravitational attractive force is completely determined by the solution to the second order vector equation \eqref{eq:SecondOrderEquation}. However, conservation of momentum and energy provide the means to reduce the problem into a system of first order scalar equations, as we now demonstrate.

We start with angular momentum. Using the polar expressions for the position \eqref{eq:r} and velocity \eqref{eq:rdot}, we can express the angular momentum as
\[
\boldsymbol L := \boldsymbol r \times m\dot{\boldsymbol r} = r\,\widehat{\boldsymbol r} \times m\left(\dot r\,\widehat{\boldsymbol r} + r\,\dot\theta\,\widehat{\boldsymbol\theta}\right) = mr^2\dot\theta\,\widehat{\boldsymbol r}\times\widehat{\boldsymbol\theta}.
\] 
This, together with \eqref{eq:CrossProd} and the conservation property imply that, under a gravitational force, the magnitude
\begin{equation}\label{eq:L}
L := |\boldsymbol L| = mr^2\dot\theta
\end{equation}
remains constant over time. Analogously, the conservation of energy implies that the quantity
\begin{equation}\label{eq:Energy}
E = \frac{1}{2}m(\dot r^2 + r^2\dot\theta^2) -\frac{GMm}{r}.
\end{equation}
is constant. These two equations involve only first derivatives of the radial and angular coordinates and for that reason are sometimes referred to as \textit{first integrals of motion}. Solving for $\dot r$ and $\dot\theta$ in the expressions above yields the first order nonlinear system of differential equations:
\begin{subequations}\label{eq:MotionEqs}
\begin{align}
\label{eq:MotionEqsA}
\dot{\theta} =\,& \frac{L}{mr^2}\,, \\[1ex]
\label{eq:MotionEqsB}
\dot{r} =\,& \sqrt{\frac{2}{m} \left(E + \frac{GMm}{r}\right) - \left(\frac{L}{mr}\right)^2}\,.
\end{align}
\end{subequations} 
If initial conditions $r(0)$ and $\theta(0)$ are given, it is possible to integrate the system \eqref{eq:MotionEqs} to obtain the time dependence for the radial and angular parameters $r$ and $\theta$ of the particle for all times. We will refer to the set of points $(r(t),\theta(t))$ that can be described by the solutions of the system \eqref{eq:MotionEqs} as \textit{orbits} or \textit{trajectories}. Alternatively, the desired initial and final positions (instead of initial position and velocity) may be prescribed. In that case, determining the orbits $(r(t),\theta(t))$ becomes something that is known as a \textit{boundary value problem} in the mathematical literature. Boundary value problems require slightly different mathematical techniques than initial value problems and, even when they are solvable, often call for additional physical constraints to avoid having multiple solutions. As we will see in Section \ref{sec:Lambert}, Lambert's problem falls into the latter category. However, before switching our attention to it, we will spend some time studying the properties of the orbits available to objects under a gravitational potential. We will then use this geometric information to avoid having to deal with differential equations when tackling Lambert's boundary value problem.

\subsection{Gravitational orbits}\label{sec:GraviatationalOrbits}
In general, under the influence of a gravitational potential the polar parameters $r$ and $\theta$ are not independent of each other, as their time behavior is constrained to obey the system of equations \eqref{eq:MotionEqs}. Since equation \eqref{eq:MotionEqsA} couples the time evolution of $r$ and $\theta$ knowledge about one of the variables will implicitly determine the other one.

The exception to this, however, is the case when $L=0$, as in that case  \eqref{eq:MotionEqsA} is independent of $r$---note that equation \eqref{eq:MotionEqsB} depends exclusively on $r$. We will study this case first. If the angular velocity $\dot\theta$ vanishes at any given time, then the angular momentum would vanish as well. Therefore, by the conservation property encoded in equation \eqref{eq:MotionEqsA}, this would imply that $\dot\theta=0$ for all times and the particle would describe a straight line at a constant angle $\theta_0 = \theta(0)$.

On the other hand, if the angular speed $\dot\theta$ is not zero initially, then the conservation of angular momentum implies that it will never vanish---as the left hand side of \eqref{eq:MotionEqsA} would then be a positive constant. In that case $r$ and $\theta$ are functionally dependent on each other and we can assume an implicit relationship of the form $r = r(\theta(t))$. Using the chain rule we obtain:
\[
\dot r = \frac{d r}{d\theta} \dot\theta.
\]
Since $\dot\theta\neq 0$ for all times, we can solve the equation above for $dr/d\theta= \dot r/\dot \theta$ and use the equations of motion \eqref{eq:MotionEqs} to obtain
\[
\frac{1}{r^2}\frac{dr}{d\theta} = \sqrt{\frac{2m}{L^2} \left(E + \frac{GMm}{r} \right) - \frac{1}{r^2}}.
\]
The reason why we kept the factor $1/r^2$ on the left hand side of the equality, is that it highlights the fact that the expression above depends on $r$  only through the function
\[
u = \frac{1}{r}.
\]
Performing this change of variables in the equation above yields
\begin{equation}\label{eq:du/dtheta}
-\frac{du}{d\theta} = \sqrt{\frac{2m}{L^2} \left(E + GMmu \right) - u^2}.
\end{equation}
The radical in the right hand side is suggestive of a trigonometric substitution, but such a substitution requires the variable $u$ to appear solely as part of a ``perfect square". We therefore complete the square inside of the square root to obtain
\[
\frac{du}{d\theta} = - \sqrt{\frac{2m}{L^2} \left(E + GMmu \right) - u^2} = -\sqrt{\frac{2mE}{L^2}+\left(\frac{GMm^2}{L^2}\right)^2 - \left(u-\frac{GMm^2}{L^2}\right)^2}.
\]
Finally, defining
\[
K := \sqrt{\frac{2mE}{L^2}+\left(\frac{GMm^2}{L^2}\right)^2} \qquad \text{ and } \qquad y:= u-\frac{GMm^2}{L^2},
\]
and realizing that ${\displaystyle \frac{du}{d\theta} = \frac{dy}{d\theta}}$, the equation \eqref{eq:du/dtheta} results in
\[
\frac{dy}{d\theta} = -\sqrt{K^2 - y^2}.
\]
This equation can be readily integrated by separation of variables yielding
\[
\cos^{-1}\left(\frac{y}{K}\right) = \theta - \theta_0.
\]
It is customary to choose the integration constant $\theta_0$ so that at initial time the particle is located at the point nearest to the focus---known as the \textit{perihelion} or \textit{periapsis}. Choosing this angle to be $\theta_0=0$ fixes an orientation of the system similar to the one depicted in Figure \ref{fig:SecondLaw}, where the axis of symmetry is horizontal and, if there are indeed two foci, the attractive center is located at the one on the right. This is the most common choice in the literature. Following this convention the equation above implies that
\[
y = K\cos\theta.
\]
Substituting back $K$ and $y$, and solving for $u$ we obtain
\[
u = \frac{GMm^2}{L^2} + \cos\theta\sqrt{\frac{2mE}{L^2}+\left(\frac{GMm^2}{L^2}\right)^2} 
= \frac{GMm^2}{L^2}\left( 1 +  \,\,\cos\theta\sqrt{\frac{2EL^2}{(GMm)^2m}+1}\right).
\]
Finally, recalling that $u=1/r$, the expression above yields 
\begin{subequations}
\begin{equation}\label{eq:randtheta}
r = \frac{p}{1 + e\cos\theta},
\end{equation}
where
\begin{equation}\label{eq:physicalexentricity}
p:= \frac{L^2}{GMm^2} \qquad \text{ and } \qquad e:= \sqrt{1+\frac{2EL^2}{(GMm)^2m}}\;\;.
\end{equation}
\end{subequations}
Equation \eqref{eq:randtheta} describes all the possible trajectories for particles \textit{with non--zero angular momentum} moving under a gravitational potential\footnote{We discussed the case of zero angular momentum at the beginning of this section, which results in straight lines connecting the initial position of the particle with the attractive center.}. The attentive reader will note that \eqref{eq:randtheta} is exactly of the form \eqref{eq:PolarConic} that we arrived at when studying conic sections. This allows us to state the following remarkable fact:
\begin{adjustwidth}{1.5cm}{0cm}\textit{
The only possible trajectories for a particle under a central gravitational potential are conic sections.
}
\end{adjustwidth}
From a purely geometric point of view, the particular conic is determined by the eccentricity. Equation \eqref{eq:physicalexentricity} connects the geometry of the problem to the physical parameters by quantifying the way in which the eccentricity $e$ depends on the particle's energy $E$ and angular momentum $L$. Concretely, expressing $p$ in terms of the angular momentum and equating \eqref{eq:randtheta} with \eqref{eq:PolarConicB}, we can connect the geometric parameters $e$ and $a$ with the physical parameters, resulting in
\begin{equation}\label{eq:Lae}
L^2 = a\left|1-e^2\right|GMm^2.
\end{equation}
We will now discuss the particular orbits and their connection with these physical parameters:\\
\begin{itemize}
\item $\boxed{\boldsymbol{L=0\,.} \text{\textbf{ Straight lines.}}}$ If the angular momentum is zero, then \eqref{eq:MotionEqsA} implies that the angle remains constant and the particle's orbit is a straight line.

\item $\boxed{\boldsymbol{L\neq0\,.}}$ If the angular momentum is different from zero, then the orbit's eccentricity $e$ is determined by the particle's initial energy
\[
E =   \underbrace{\frac{1}{2}m(\dot r_0^2 + r_0^2\dot\theta_0^2)}_{\text{\small Kinetic}} \,\, \underbrace{- \,\,\frac{GMm}{r_0}}_{\text{\scriptsize Potential}},
\]
where $r_0, \theta_0,\dot r_0$ and $\dot\theta_0$ denote respectively the initial radial and angular positions and velocities. The energy can be negative, zero or positive, which leads to three possible geometric scenarios:\\
\begin{enumerate} 

\item[(a)] $\boxed{E < 0\,. \, \text{\textbf{Closed orbits}}.}$ When the attractive potential is larger than the particle's kinetic energy, the total energy of the system will be negative. In this case the particle won't have enough energy to escape the attractive potential and will remain ``trapped" in the sense that its distance $r$ to the attractive potential will be bounded for all times. 

Going back to the equation \eqref{eq:physicalexentricity} for the eccentricity, we note that the only term inside the radical that is not necessarily positive is the total energy $E$. Since the eccentricity must be a real number, the admissible values for the energy are limited by the condition
\[
1 + \frac{2EL^2}{(GMm)^2m} \geq 0.
\]
Thus, the admissible negative values for the energy lie in the range
\[
-\frac{(GMm)^2m}{2L^2} \leq E < 0.
\]
We distinguish two cases:\\
\begin{enumerate}
\item[(i)] $\boxed{e =0\,.\, \text{\textbf{Circular orbits}}.}$ When substituted into Equation \eqref{eq:physicalexentricity}, the minimum admissible energy 
\[
E = -\frac{(GMm)^2m}{2L^2},
\]
yields and eccentricity $e=0$, which corresponds to a circular orbit. To determine the radius of the circle, we let $e=0$ in Equation \eqref{eq:randtheta} and conclude that the distance between the particle and the attractive center (i.e. the radius of the orbit) is
\[
r = \frac{L^2}{GMm^2}.
\]
\item[(ii)] $\boxed{0<e<1\,.\, \text{\textbf{Elliptic orbits}}.}$ For values of the energy in the interval
\[
-\frac{(GMm)^2m}{2L^2} < E < 0
\]
the eccentricity of the orbit remains strictly inside the interval $(0,1)$, which corresponds to an ellipse. The minimum and maximum distances prescribed by \eqref{eq:randtheta} are
\[
r_{\text{max}} = \frac{L^2}{GMm^2(1-e)} \qquad \text{ and } \qquad r_{\text{min}} = \frac{L^2}{GMm^2(1+e)}.
\]
From these, we can express the semi--major axis of the ellipse in terms of physical parameters as
\begin{equation}\label{eq:aphysical}
a = \tfrac{1}{2}(r_{\text{max}} + r_{\text{min}}) = \frac{L^2}{GMm^2 (1-e^2)}.
\end{equation}
\end{enumerate}

The expressions for the radius of circular orbits and the semi--major axis of elliptic orbits that we just derived, confirm that a particle with no angular momentum necessarily collapses into the attractive center. They also reveal that, for a fixed initial angular momentum, more massive bodies  (i.e. those for which the product $Mm^2$ is larger) are bound to remain closer to the attractive center.

\item[(b)]  $\boxed{E \geq 0\,.\, \text{\textbf{Open orbits}}.}$ When the particle's kinetic energy is enough to exactly counteract the attractive potential or exceeds it, the particle will be able to ``escape" the attraction in the sense that its distance to the attractive center will eventually grow unbounded. We infer this from the fact that $e^{-1}<1$ and therefore, form \eqref{eq:randtheta}, we have that
\[
r_{\text{max}} = \lim_{\theta\to \pm\cos^{-1}(-e^{-1})}\frac{p}{1 + e\cos\theta} = \infty.
\]
\
The angles $\theta_{\pm\infty}:=  \pm\cos^{-1}(-e^{-1})$ appearing in the limit above determine oblique asymptotic lines that the particle cannot cross. As the particle drifts away, the conservation condition dictates that
\begin{equation}\label{eq:EnergyAtInfty}
E = \lim_{r\to\infty} \left(\frac{1}{2}m(\dot r^2 + r^2\dot\theta^2)-\frac{GMm}{r}\right) = \lim_{r\to\infty}\left(\frac{1}{2}m(\dot r^2 + r^2\dot\theta^2)\right) < \infty.
\end{equation}
For this to hold, we must have that 
\[
\lim_{r\to\infty} \dot\theta = 0 \qquad  \text{ and } \qquad  \lim_{r\to\infty} \dot r = \dot r_{\infty} < \infty.
\]
In fact, the condition on the angular velocity is even stronger, requiring
\[
{\displaystyle \lim_{1/r\to 0} \frac{\dot\theta}{(1/r)} = 0}.
\]
Hence, as a particle drifts away to infinity, its angular velocity decays rapidly and indeed vanishes before the angular coordinate of the particle can reach the angles $\theta_{\pm\infty}$. Particles in this situation will describe open trajectories. We distinguish between two cases:\\

\begin{enumerate}

\item[(iii)]  $\boxed{e = 1 \,.\, \text{\textbf{Parabolic orbits.}}}$ If the potential term equals the kinetic term, the total energy will be $E=0$. Equation \eqref{eq:physicalexentricity} implies that $e=1$ and the particle follows a parabolic orbit. Comparing \eqref{eq:randtheta} and \eqref{eq:PolarConicA} it follows that the distance between the directrix of the parabola and the attractive center is
\[
r_{\text{min}}= |FL| = p = \frac{L^2}{GMm^2},
\]
while, as the angle $\theta$ approaches  $ \pi$ or $-\pi$, equation \eqref{eq:randtheta} implies that $r\to\infty.$ We note that as $r\to\infty$ the potential energy vanishes. The limit condition \eqref{eq:EnergyAtInfty} and the fact that $E=0$ imply that the residual speed at infinity for a particle on a parabolic trajectory is $\dot r_{\infty} = 0$. 

\item[(iv)] $\boxed{e > 1 \,.\, \text{\textbf{Hyperbolic orbits.}}}$ Finally, when the kinetic energy exceeds the potential energy the eccentricity of the trajectory will be greater than one, and the particle will follow a hyperbolic arc. In this case the minimum distance to the attractive center will be given by
\[
r_{\text{min}} = \frac{L^2}{GMm^2(1+e)}.
\]
The excess in kinetic energy allows these particles to ``reach infinity" with a non--zero radial velocity given by $\dot r_{\infty} = \sqrt{2E/m}$.
\end{enumerate}
\end{enumerate}
\end{itemize}

In the elliptic and hyperbolic cases, the numerators from \eqref{eq:PolarConicB} and  \eqref{eq:randtheta} can be equated, and the form of $e$ in \eqref{eq:physicalexentricity} can be used to obtain the expression for the semi--major axis $a$ in terms of the physical parameters
\begin{equation}\label{eq:aenergy}
a = \frac{GMm}{2|E|},
\end{equation}
which in the limiting case $E\to 0$ is consistent with the fact that the length of the axis of symmetry of a parabola is infinite.

\subsection{Kepler's laws of planetary motion}

In their modern version, Kepler's laws are usually stated as slight variations of the following \cite{Morin2012,Scheck2010}:
\textit{\begin{enumerate}
\item The planets move in elliptical orbits with the Sun at one focus.
\item The radius vector from the Sun to the planets sweeps out equal areas in
equal times.
\item The square of the period of an orbit, $\mathcal T$, is proportional to the cube of the
semi--major axis length, $a$. More precisely,
\[
\mathcal T^2 = \frac{4\pi^2a^3}{GM},
\]
where $M$ is the mass of the Sun.
\end{enumerate} }
We will now use the analytical tools developed in the previous section to prove Kepler's statements.


\textbf{\textit{First law.}} \textit{The planets move in elliptical orbits with the Sun at one focus.}

The modern--day proof of this claim is, of course, the detailed analytic process leading to equation \eqref{eq:randtheta}, combined with the empirical observation that the distance between the planets and the Sun is bounded (which discards hyperbolas, parabolas and straight lines). It is important to remark that, in full rigor, this statement is only true for the interaction between the Sun and \textit{a single} planet.


\textbf{\textit{Second law.}} \textit{The radius vector from the Sun to the planets sweeps out equal areas in equal times.}

\begin{figure}
\centering
\includegraphics[width = 0.33\linewidth]{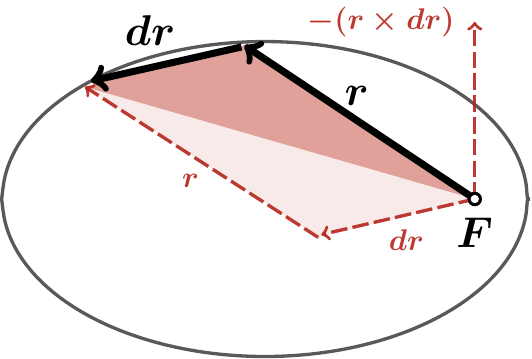} \hspace{2cm}
\includegraphics[width = 0.33\linewidth]{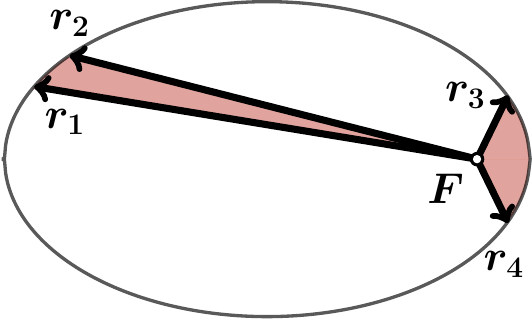}
\caption{The area of the parallelogram determined by the vectors $\boldsymbol r$ and $\boldsymbol{dr}$ (shaded) is given by the norm $|\boldsymbol{r\times dr}|$. As the position vector $\boldsymbol r$ undergoes a small displacement $\boldsymbol{dr}$ along the elliptic arc, the area of the infinitesimal elliptic sector it traverses (dark shade) is given by $\tfrac{1}{2}|\boldsymbol{r\times dr}|$. Right: The time required to travel between $\boldsymbol{r_1}$ and $\boldsymbol{r_2}$ is the same as that required to travel between $\boldsymbol{r_2}$ and $\boldsymbol{r_4}$.}\label{fig:SecondLaw}
\end{figure}

This fact is a consequence of the conservation of angular momentum, as we now show. Consider a particle with position vector $\boldsymbol r$ traveling along an elliptical trajectory. After experiencing an infinitesimal displacement $\boldsymbol{dr}$, its position vector will sweep an infinitesimal elliptic sector as depicted in Figure \ref{fig:SecondLaw}. From elementary vector algebra, the area $dA$ of the infinitesimal sector will be given by
\[
dA = \tfrac{1}{2}|\boldsymbol{r \times dr}|.
\]
A computation completely analogous to the one leading to Equation \eqref{eq:rdot}, shows that
\[
\boldsymbol{dr} = dr\,\widehat{\boldsymbol r} + rd\theta\,\widehat{\boldsymbol \theta}.
\]
Using this and recalling that the unit vectors $\widehat{\boldsymbol r}$ and $\widehat{\boldsymbol \theta}$ are orthogonal we obtain
\[
dA = \tfrac{1}{2}|r\,\widehat{\boldsymbol r} \times (dr\,\widehat{\boldsymbol r} + rd\theta\,\widehat{\boldsymbol \theta}) | = \tfrac{1}{2}|r \, \widehat{\boldsymbol r} \times rd\theta \,\widehat{\boldsymbol \theta} | = \tfrac{1}{2} r^2 d\theta.
\]
Hence, the area traversed by the particle between the times $t_0$ and $t$ is given by
\begin{equation}\label{eq:Kepler3}
A = \; \tfrac{1}{2}\int_{\theta(t_0)}^{\theta(t)}r^2 d\theta \; =\;  \tfrac{1}{2}\int_{t_0}^{t}r^2 \dot\theta ds \underbrace{=}_{\text{\scriptsize Conservation}} \tfrac{1}{2}\int_{t_0}^{t}\tfrac{L}{m}\,ds \; = \; \tfrac{L}{2m}(t-t_0),
\end{equation}
where in the third equality we used the fact that $|\boldsymbol L| = L =mr^2\dot\theta$ is constant with respect to time. Therefore, the area swept depends only on the length of the time interval elapsed, which is equivalent to Kepler's statement.

\textbf{\textit{Third law}.} \textit{The square of the period of an orbit is proportional to the cube of the semi--major-axis length.}

Setting the time span $t-t_0$ equal to one period of revolution, $\mathcal T$, in \eqref{eq:Kepler3} and recalling that the area of an ellipse is given by $\pi ab$ (where $a$ and $b$ are the lengths of the semi--major and semi--minor axes of the ellipse), we obtain
\begin{equation}\label{eq:piab}
\pi ab = \frac{L\mathcal T}{2m}.
\end{equation}
We can then use the last equality in \eqref{eq:OF} to express $a = \frac{e|FL|}{|1-e^2|} $ and substitute this into \eqref{eq:SemiMinor} yielding
\[
b = |FL|\sqrt{\frac{e^2}{|1-e^2|}} = a \sqrt{|1-e^2|}.
\] 
From here, we use \eqref{eq:aphysical} to express $\sqrt{1-e^2} = \frac{L}{m}\sqrt{\frac{1}{GMa}}$ and obtain
\[
b = \frac{L}{m}\sqrt{\frac{a}{GM}}
\]
which, upon substitution, turns the expression \eqref{eq:piab} into
\[
\frac{\pi L}{m}\sqrt{\frac{a^3}{GM}} = \frac{L\mathcal T}{2m} . 
\]
Finally, squaring both sides and solving for $\mathcal T$ leads to
\begin{equation}\label{eq:K3}
\mathcal T^2 = \frac{4\pi^2 a^3}{GM},
\end{equation}
as desired.  
\subsection{Brief historical digression}

Kepler's celebrated laws were published between 1609 and 1619 and had a tremendous impact on the astronomical community of the early $17^{th}$ century and beyond. They are the result of decades of careful measurements of the planets's locations made by Tycho Brahe, and then of several more years of detailed mathematical analysis of the measurements (mostly of the orbit of Mars) by Kepler himself \cite{Wilson1972}. They are a testament of Brahe's meticulous measurements (all of them made \textit{with the naked eye} \cite{Tabak2011}, as the telescope would not be around until 1609) as well as of Kepler's mathematical and computational prowess and scientific integrity. Although by Kepler's time the Heliocentric theory of the universe was starting to show serious cracks, the notion that planets followed circular orbits was still widely accepted as true. Kepler estimated the eccentricity of Mars's orbit to be $e\approx 0.0926$ \cite{Xavier}---which is not too far from that of a circle. Due to the coarseness of most contemporary measurements, Kepler could have easily attributed the discrepancy to measurement or computational error, however he trusted Brahe's measurements and his own computations. Years of wrestling with the data had convinced him that the circle was not the correct answer and he pushed forward his \textit{elliptic} discovery, even if it clearly went against the widely accepted theories.

As originally stated by Kepler, his laws were purely kinematic descriptions that did not offer an underlying theory explaining the origin of the movement \cite{Hockey2014,Katz2008}. Due to the lack of a guiding principle leading Kepler's analysis of empirical data, it can be argued that his conclusions were the product of serendipity: Brahe's measurements were precise enough to show a discrepancy from a circular orbit, but not precise enough to reveal the perturbations due to the gravitational fields of other planets, which would have led Kepler astray \cite{Stillwell2010}. A complete dynamic explanation of planetary orbits would have to wait until Newton's inverse square law, while the powerful and elegant analytical treatment that we have used above would require a further refinement of Newton's calculus. The arguments and tools that we used in the previous section to obtain the equation of the orbits can be traced back to Euler and Lagrange in the mid to late $18^{th.}$ century \cite{Stillwell2010}.

What we now call Kepler's first law appeared in 1609, in his book \textit{Astronomia nova} \cite{Kepler1609}, where he carefully describes the lengthy thought process that led him to the conclusion that Mars's orbit is elliptic, and states that:
\begin{adjustwidth}{3cm}{0cm}\textit{
``. . . the orbit of the planet is not a circle, but comes in gradually on both sides and
returns again to the circle’s distance at perigee. They are all accustomed to call the shape of this sort of path ‘oval’." \cite{Kepler1609,Linton2004}
}
\end{adjustwidth}
Kepler in fact derived an equation that is almost identical to the center--based equation for an ellipse, but using a different angular parameter that became known as the \textit{eccentric anomaly} (depicted in Figure \ref{fig:TrueAndEccentricAnomaly}, where it is denoted as $\phi$) \cite{Linton2004}. After exhaustively working with the data for Mars and concluding that the ellipse was the only possible explanation, Kepler barely verified the data for other planets before generalizing the conclusion based on the conviction that ``the harmony of nature demanded that all \textit{have similar habits}" \cite{Burton2010}.  Regarding the location of the Sun at a focus of the ellipse, Linton \cite{Linton2004} points out that: ``nowhere in the main body of the \textit{New Astronomy} is the word ‘focus’ mentioned, and it was only later in his \textit{Epitome of Copernican Astronomy} that Kepler emphasized this aspect of planetary orbits". The \textit{Epitome of Copernican Astronomy}, originally entitled ``Epitome Astronomiae {C}opernicanae" \cite{Kepler1995} was intended to be a textbook in astronomy and was published in seven parts between 1618 and 1621 \cite{Caspar2012,Rothman2020}.  

As for the second law, Kepler derived it as an approximation, in an attempt to find a proxy for the time elapsed while a planet describes an arc. His derivation uses an argument that can be considered incipient integral calculus: as described by Katz \cite{Katz2008}, ``Kepler then argued that the total time required to pass over a finite arc [...] could be thought of as the sum of the radius vectors making up that part of the circle, or as the area swept by the radius vector". Kepler was no stranger to ``infinitesimal" arguments of that sort, which come remarkably close to integration. During his career he would use similar reasoning to approximate the area of a circle by considering the sum of infinitely many triangles with vertex on the center of the circle and sides along the circumference, as well as to calculate the surface area of a sphere by a similar argument involving infinitesimal cones \cite{Burton2010}. He famously went on to extend these techniques to calculating the volume of different solids of revolution and wine barrels \cite{Kepler2018}.

Going back to the \textit{equal areas} law, Kepler discovered it sometime before 1605 and used it extensively in the calculations that would eventually lead to the first law, as he himself describes in his 1609 \textit{New Astronomy}. However, his thoughts on this rule evolved over time and he eventually concluded that it was not an approximation, but a true statement. He would end providing a geometric proof that is almost correct by modern standards in 1621, as part of volume V of his \textit{Epitome of Copernican Astronomy} \cite{Davis2003}.

The fifth volume of the book \textit{Harmonici mundi} (harmony of the world), published in 1619 \cite{Kepler1619}, contains---without much fanfare---the statement that would become known as his third law. Kepler discovered this proportionality relationship around 1618 while attempting to find a connection between the movement of the planets and musical harmonies (a problem that would obsess him for his entire adult life). The discovery was also based on the analysis of empirical data, but Kepler did not consider it important enough to provide a table of the measurements in the text. Later on, in the \textit{Epitome}, he would attempt---incorrectly---to explain the origin of the third law by assuming that the volumes of the planets were proportional to their distance from the Sun, and their densities decayed inversely as the square root of their distance to the Sun \cite{Linton2004}.

\section{Lambert's problem}\label{sec:Lambert}

As we concluded from the argument developed in Section \ref{sec:GraviatationalOrbits}, the only admissible trajectories for an object moving solely under the influence of a gravitational potential are conic sections. We will refer to movement along these trajectories as ``free falling"  since, once the particle is placed on a given orbit, conservation of angular momentum takes over and no additional energy is required to set it in motion. As is evident from equation \eqref{eq:physicalexentricity}, considering that the masses of the particle and the attractive center are fixed and given, the particular conic section will be determined by the particle's energy---or equivalently its velocity. 

Lambert's problem can be expressed as:
\begin{problem}[Lambert's Problem]\label{prob:Lambert}
Given a particle located at the position $\boldsymbol r_1$ and a target position $\boldsymbol r_2$, determine the energy required to place the particle on a conical orbit connecting the two points, and such that the trip's duration is $\Delta t$ while completing $Q$ revolutions before arriving at $\boldsymbol r_2$.
\end{problem}
From the mathematical point of view, this question becomes a two--point boundary value problem involving the equations of motion \eqref{eq:MotionEqs} (or equivalently equation \eqref{eq:SecondOrderEquation}). Boundary value problems differ from initial value problems in the sense that, instead of knowing the initial position and velocity, $\boldsymbol r$ and $\boldsymbol{\dot r}_0$,
of the particle (or alternatively its energy and angular momentum), the initial and final positions $\boldsymbol r_0$ and $\boldsymbol r_f$ are prescribed. This condition changes substantially the mathematical treatment of the problem. One key difference is that boundary value problems tend to have multiple solutions and therefore need supplemental physical information to elucidate the correct answer. Intuitively, a particle could travel along the same---closed---conic section either clockwise or counter--clockwise and complete multiple cycles before arriving at the final position. As we shall soon see, in our case the additional information will pertain to the number of revolutions that the particle can complete before arriving at the target endpoint. We will appeal to the geometric analysis of the possible trajectories that we developed in the previous section to sidestep the problem of \textit{integrating} the boundary value problem.

In what follows, our analysis will be focused only on particles that follow elliptical trajectories. We will consider that the ellipse has a semi--major axis of length $a$, a semi--minor axis of length $b$, eccentricity $0\leq e<1$ (understanding the circle as the limiting case of ellipses as $e\to 0$), and the attracting center is located at the focus $F_1$ as depicted in Figure \ref{fig:TrueAndEccentricAnomaly}.  Derivations following similar arguments for other conical orbits can be found in \cite{Grossman1996,Izzo2014,LaBl1969,LaBlDe1966}---although the exposition in those sources is less detailed. 

\subsection{Eccentric anomaly}

\begin{figure}
\centering
\includegraphics[height = 0.3\linewidth]{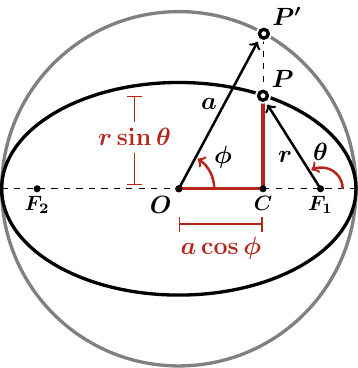} \hspace{.175\linewidth}
\includegraphics[height = 0.3\linewidth]{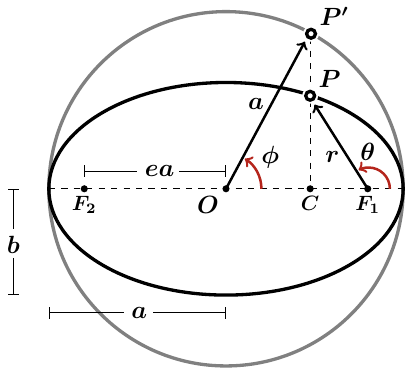} 
\caption{Left: The horizontal coordinate of a point $P$ in the ellipse can be expressed in terms of the eccentric anomaly $\phi$, while the vertical coordinate can be expressed in terms of the true anomaly $\theta$. The segments are marked in red in the diagram. The expressions for the two coordinates can be combined through the Cartesian equation for an ellipse to produce a relation between the two anomalies. Right: The position $P$ of a point in an ellipse, centered at the origin with semi--major axis $a$, can be described by its distance $r$ to one of the foci and the angle $\phi$ subtended by the radius joining the origin and the vertical projection $P^\prime$ of the point $P$ onto an auxiliary circle of radius $a$ centered at the origin. The angle $\phi$ is known as the \textit{eccentric anomaly}.}\label{fig:TrueAndEccentricAnomaly}
\end{figure}

Up to this point we have described the position $P$ of a particle by its distance, $r$, to the attractive center located at the focus, and the angle $\theta$ subtended by the vector connecting $P$ to $F_1$, i.e. the \textit{true anomaly}. However, in some cases it is convenient to characterize the position as if the particle were traversing a circle. This is achieved by inscribing the elliptical trajectory inside of an auxiliary circle of radius $a$, and projecting the point $P$ vertically (in the same up/down direction as the vector connecting $F_1$ to $P$) onto the auxiliary circle as depicted in Figure \ref{fig:TrueAndEccentricAnomaly}. We denote by $P^\prime$ and $C$ the points where the vertical line passing through $P$ intersects the auxiliary circle and the horizontal line connecting the foci, respectively. The angle $\phi$ formed between the horizontal axis and the vector connecting the origin $O$ and the point $P^\prime$ is known as the \textit{eccentric anomaly}.

We first note that, as depicted in the right panel of Figure \ref{fig:TrueAndEccentricAnomaly}  (where both segments are highlighted in red), the horizontal coordinate of a point $P$ on the ellipse can be written in terms of the eccentric anomaly as $a\cos\phi$, while its vertical component is easily expressed in terms of the true anomaly as $r\sin\theta$. Hence a point in the ellipse has coordinates given by 
\begin{equation}\label{eq:xy}
(x,y) = (a\cos\phi, r\sin\theta).
\end{equation}
Recalling that the lengths of the semi--major and semi--minor axes of the ellipse are $a$ and $b$, respectively, and that the point $P=(a\cos\phi, r\sin\theta)$ must satisfy the Cartesian equation of the ellipse
\[
1 = \frac{x^2}{a^2} + \frac{y^2}{b^2} = \frac{a^2\cos^2\phi}{a^2} + \frac{r^2\sin^2\theta}{b^2} 
\]
we obtain
\begin{equation}\label{eq:cosphi}
r\sin\theta = b\sin\phi. 
\end{equation}
This relation, which together with \eqref{eq:xy} amounts to the familiar parametrization of an ellipse in terms of its central angle in the form $(a\cos\phi,b\sin\phi)$, will prove useful soon.

Our next goal is to express the position $P$ in terms of the radius $r$ and the eccentric anomaly $\phi$. We start by observing that the length of the segment $CF_1$ appearing in Figure \ref{fig:TrueAndEccentricAnomaly} can be expressed in two different ways. On the one hand, it is one of the legs of the right triangle $CF_1P$, and therefore 
\[
|CF_1|= -r\cos\theta.
\]
On the other hand, the length can be computed by subtracting the length of the segment $OC$ from that of the segment $OF_1$ (i.e. the focal length), which yields
\[
|CF_1| = |OF_1|-|OC| = ae - a\cos\phi. 
\]
Equating the two expressions above and solving for $\cos\theta$ results in
\[
\cos\theta = \frac{a(\cos\phi-e)}{r}.
\]
From the focal representation of the ellipse \eqref{eq:PolarConicB}, we can obtain
\[
\cos\theta = \frac{1}{e}\left(\frac{a(1-e^2)}{r} - 1\right).
\]
The true anomaly, $\theta,$ can be eliminated by equating the last two expressions. Solving then for $r$ in the ensuing equality yields the following relationship between $r$ and the eccentric anomaly $\phi$:
\begin{equation}\label{eq:rphi}
r = a(1-e\cos\phi).
\end{equation}
This equation is the equivalent of the polar equation of the ellipse, when the position of the point along the ellipse is identified by its distance to the focus, $r$, and its eccentric anomaly $\phi$.

\subsection{Kepler's equation}

Thus far, we have obtained two analytic expressions for the position of a particle on an ellipse: the focal equation of the ellipse \eqref{eq:PolarConicB} and the eccentric equation \eqref{eq:rphi}, both of which relate the distance to the focus $r$ to a different angular parameter. Equating these two expressions results in
\[
\frac{a(1-e^2)}{1+e\cos\theta} = r = a(1-e\cos\phi).
\] 
Differentiating the expression above with respect to time yields
\[
\frac{ae(1-e^2)\dot\theta\sin\theta}{(1+e\cos\theta)^2} = \dot r = ae\dot\phi\sin\phi.
\]
Solving for $\dot\phi$ and simplifying we obtain
\begin{alignat*}{6}
\dot\phi =\,& \frac{(1-e^2)}{(1+e\cos\theta)^2}\cdot\frac{\dot\theta\sin\theta}{\sin\phi} \qquad\qquad && \\[1ex]
=\,& \frac{r^2}{a^2(1-e^2)}\cdot\frac{\dot\theta\sin\theta}{\sin\phi} && (\text{From }\eqref{eq:PolarConicB}) \\[1ex]
=\,& \frac{1}{a^2(1-e^2)}\cdot\frac{L}{m}\cdot\frac{\sin\theta}{\sin\phi} && (\text{From }\eqref{eq:L} ) \\[1ex]
=\,& \frac{1}{a\sqrt{1-e^2}}\cdot\frac{L}{m}\cdot\frac{\sin\theta}{b\sin\phi} && (\text{From }\eqref{eq:AandB} ) \\[1ex]
=\,& \frac{1}{a\sqrt{1-e^2}}\cdot\frac{L}{m}\cdot\frac{\sin\theta}{r\sin\theta} && (\text{From }\eqref{eq:cosphi} ) \\[1ex]
=\,& \left( \frac{m}{L}\,\sqrt{\frac{GM}{a}}\right)\frac{L}{m}\cdot\frac{1}{r} \qquad&& (\text{From }\eqref{eq:Lae} ) \\[1ex]
=\,& \frac{a}{r}\sqrt{\frac{GM}{a^3}} \; = \; \frac{a}{r}\cdot\frac{2\pi}{\mathcal T} && (\text{From }\eqref{eq:K3} ) \\[1ex]
=\,& \frac{1}{(1-e\cos\phi)}\cdot\frac{2\pi}{\mathcal T} && (\text{From }\eqref{eq:rphi} ).
\end{alignat*}
The ratio $2\pi/\mathcal{T}$ is the average angular speed of the particle undergoing a full orbit over a period $T$; it is referred to as the \textit{mean motion} and sometimes denoted by the letter $n$. We have thus obtained the relation
\[
(1-e\cos\phi)\dot\phi = \frac{2\pi}{\mathcal{T}},
\]
which can be readily integrated with respect to time to obtain
\begin{equation}\label{eq:Kepler}
\phi - e\sin\phi = \frac{2\pi}{\mathcal{T}}(t - t_0),
\end{equation}
where the integration constant $-2\pi t_0/\mathcal{T}$ follows from the convention on setting the initial position at the perihelion (or periapsis), where the angle $\phi$ vanishes. This equation is known as Kepler's equation.

The quantity $2\pi(t-t_0)/\mathcal T$ appearing on the right hand side of \eqref{eq:Kepler} has \textit{angular} units (recall that $2\pi=360^\circ$ is the angular measure of a circle in radians) and is called the \textit{mean anomaly}. It can be interpreted as the angular position of a fictitious body that moves with constant angular velocity $2\pi/\mathcal T$ around a circular orbit of radius $a$. It is sometimes denoted by $M$ in the literature. To avoid confusing it with the mass of the  attractive center, we will not follow this convention in the text and will write out the full quotient explicitly instead. The fact that the mean anomaly is an angular measure that varies linearly with time enables the use of time intervals as proxies for angular displacement. If the \textit{actual} angular displacement $\phi$ from the center of the ellipse is sought for, it can be obtained from the mean anomaly by solving Kepler's equation. The angular displacement associated with the focus--centered polar coordinates $(r,\theta)$ can be obtained then from the eccentric anomaly from equation \eqref{eq:cosphi}.

\subsection{Lambert's system}

\begin{figure}[tb]
\centering
\includegraphics[height = 0.3\linewidth]{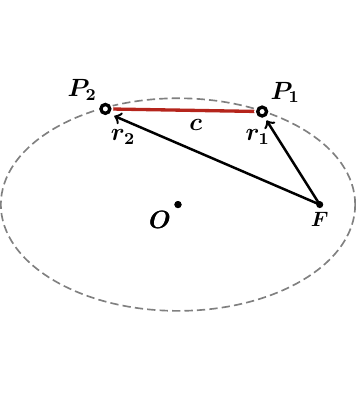} \hspace{.175\linewidth}
\includegraphics[height = 0.3\linewidth]{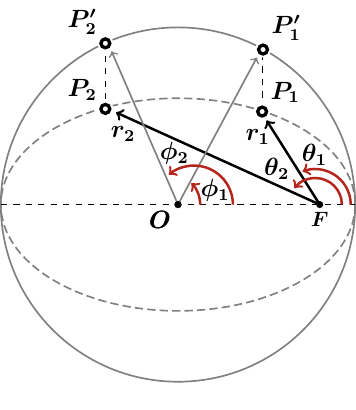} 
\caption{Left: Since the location of the attractive center $F$ and the initial and final points $P_1$ and $P_2$ are known, the lengths of the position vectors $r_1,r_2$ and the chord $c:=r_2-r_1$ can be computed. The goal is to determine the unknown ellipse (dashed line) connecting the points and such that the travel time equals a prescribed value. Right: The true anomaly, denoted by $\theta$ and eccentric anomaly, denoted by $\phi$, are angular descriptors of the points $P_1$ and $P_2$.}\label{fig:TransferGeometry}
\end{figure}

Let us recall that our goal is to ascertain the energy required to place a vessel on an elliptic orbit with the attractive center located at one of its foci and passes through two given points $P_1$ and $P_2$. In geometric terms, we must find the length of the ellipse's semi--major axis $a$, the value of the eccentricity $e$, and the angle between the major axis and the horizontal line. Once these geometric parameters are known, the energy $E$ can be recovered from Kepler's first law using equation \eqref{eq:aenergy}.

In this section, we will take advantage of Kepler's equation \eqref{eq:Kepler}, the eccentric description of the ellipse \eqref{eq:rphi}, and a simple geometric observation regarding the distance between $P_1$ and $P_2$ (i.e. the length of the chord $c:=r_2-r_1$), to obtain a system of equations that will ultimately provide us with the desired parameters. We will be making use of the following trigonometric identities, which can be easily derived from those of the sine and cosine of sums/differences of angles:
\begin{subequations}\label{eq:TrigIDs}
\begin{align}
\label{eq:TrigIDsA}
\cos\phi_1 - \cos\phi_2 &= 2 \sin\left(\!\frac{\phi_1 + \phi_2}{2}\!\right)\sin\left(\!\frac{\phi_2 - \phi_1}{2}\!\right),\\
\label{eq:TrigIDsB}
\sin\phi_1 - \sin\phi_2 &= -2 \cos\left(\!\frac{\phi_1 + \phi_2}{2}\!\right) \sin\left(\!\frac{\phi_2 - \phi_1}{2}\!\right),\\
\label{eq:TrigIDsC}
\cos\phi_1 + \cos\phi_2 &= 2\cos\left(\!\frac{\phi_1 + \phi_2}{2} \!\right) \cos\left(\!\frac{\phi_2 - \phi_1}{2}\!\right).
\end{align}
\end{subequations}

We start by noting that---since the locations of the attractive center and initial and final points are given---the lengths of the position vectors and the distance $c$ between $P_1$ and $P_2$ (as depicted in the left panel of Figure \ref{fig:TransferGeometry}) are all known. Therefore, using the eccentric description of the points 
\[
P_1 = (a\cos\phi_1, b\sin\phi_1) \qquad \text{ and } \qquad P_2 = (a\cos\phi_2, b\sin\phi_2)
\]
depicted in the right panel of Figure \ref{fig:TransferGeometry}, we can relate $c$ to the length $a$ and the eccentric anomalies $\phi_1$ and $\phi_2$ as follows:
\begin{subequations}\label{eq:Lambert}
\begin{alignat}{10}
\nonumber
c^2 &= |P_1 - P_2|^2 &&\\
\nonumber
&= |(a(\cos\phi_1-\cos\phi_2), b(\sin\phi_1-\sin\phi_2))|^2 &&\\
\nonumber
&= a^2\left((\cos\phi_1 - \cos\phi_2)^2 + (1-e^2)(\sin\phi_1 - \sin\phi_2)^2\right) \qquad\quad&& \text{\small (From \eqref{eq:AandB})}\\
\label{eq:LambertA}
&= 4a^2 \left( 1 - e^2 \cos^2\left( \frac{\phi_1 + \phi_2}{2} \right) \right) \sin^2\left( \frac{\phi_2 - \phi_1}{2} \right) &&\text{\small (From \eqref{eq:TrigIDsA} and \eqref{eq:TrigIDsB})}.
\end{alignat}
The equation above, that uses the known magnitude of the difference $|r_2-r_1|= c$, will be the first equation of our system. We now use the eccentric representation \eqref{eq:rphi} to write the radial distance in terms of the eccentric anomaly for the initial and final points $P_1$ and $P_2$ as
\[
r_1 = a(1-e\cos \phi_1) \qquad \text{ and } \qquad r_2 = a(1- e\cos \phi_2).
\] 
Adding these two equations and using \eqref{eq:TrigIDsC} we obtain
\begin{equation}\label{eq:LambertB}
    r_1 + r_2 = 2a \left(1 - e\cos\left( \frac{\phi_1 + \phi_2}{2} \right) \cos\left( \frac{\phi_2 - \phi_1}{2} \right)\right). 
\end{equation}
This expression relates the value of the sum $r_1+r_2$ to the unknown geometric parameters, and will constitute the second equation of our system. Finally, from Kepler's equation \eqref{eq:Kepler} it follows that at the times $t_1$ and $t_2$ when the vessel is located at the points $P_1$ and $P_2$ respectively, we have
\[
\phi_1 - e\sin\phi_1 = \frac{2\pi}{\mathcal{T}}(t_1 - t_0), \qquad \text{ and } \qquad \phi_2 - e\sin\phi_2 = \frac{2\pi}{\mathcal{T}}(t_2 - t_0).
\]
Subtracting the last two equations and using \eqref{eq:TrigIDsB} leads to
\[
\frac{2\pi}{\mathcal T}(t_2 - t_1) = \phi_2 - \phi_1 -2e\cos\left(\!\frac{\phi_1 + \phi_2}{2}\!\right)\sin\left(\!\frac{\phi_2 - \phi_1}{2}\!\right).
\]
Finally, we make use of Kepler's third law \eqref{eq:Kepler3} to express the period $\mathcal T$ in terms of the semi--major axis and rewrite the previous equation in the form:
\begin{equation}\label{eq:LambertC}
\sqrt{\frac{GM}{a^{3}}}(t_2-t_1) = \phi_2 - \phi_1 -2 e\cos\left(\!\frac{\phi_1 + \phi_2}{2}\!\right)\sin\left(\!\frac{\phi_2 - \phi_1}{2}\!\right).
\end{equation}
\end{subequations}
Equations \eqref{eq:LambertA}, \eqref{eq:LambertB}  and \eqref{eq:LambertC}, which sometimes are known collectively as Lambert's system, relate the geometric unknowns $a$, $e$, $\phi_1$, and $\phi_2$ to the known quantities $c = |r_2-r_1|$ and $r_1+r_2$. The time interval $t_2-t_1$ appearing in the left-hand side of \eqref{eq:LambertC}, although not known \textit{a priori}, can be prescribed as part of the problem data. These equations will be the starting point to our determination of the transfer ellipse. 

\subsection{Lagrange's solution}

A closer inspection of system \eqref{eq:Lambert} would seem to indicate that it contains one too many unknowns, as there are three equations and four unknown quantities: $a$, $e$, $\phi_1$, and $\phi_2$. However, Lagrange observed that the angles $\phi_1$ and $\phi_2$, along with the eccentricity $e$, appear in the system only in two particular combinations:
\[
\phi_2-\phi_1  \qquad \text{ and } \qquad e\cos\left(\frac{\phi_1+\phi_2}{2}\right).
\]
One would then be tempted to perform a na\"ive change of variables and proceed to attempt a solution of the system \eqref{eq:Lambert} in terms of the three unknowns $a$, $\phi_2-\phi_1$ and $e\cos\left(\tfrac{1}{2}(\phi_1+\phi_2)\right)$. However Lagrange's insight went one step further and he realized that by introducing two auxiliary variables $\alpha$ and $\beta$ such that
\begin{subequations}\label{eq:alphabeta}
\begin{alignat}{6}
\label{eq:alphabetaA}
\cos\left(\!\frac{\alpha+\beta}{2}\!\right) &= e\cos\left(\!\frac{\phi_1+\phi_2}{2}\!\right), \qquad \qquad&& 0\leq \alpha + \beta < 2\pi,\\
\label{eq:alphabetaB}
\alpha - \beta &= \phi_2-\phi_1 -2\pi Q,&& 0 \leq \alpha - \beta < 2\pi,
\end{alignat}
\end{subequations}
the system \eqref{eq:Lambert} takes a much simpler form. In the expression above, $Q$ is the number of orbital cycles completed by the vessel in the given time interval $t_2-t_1$ and the constraints on the sum a difference of the auxiliary variables account for the periodicity of equation\eqref{eq:alphabetaA}.

Performing the substitutions \eqref{eq:alphabeta} in the system \eqref{eq:Lambert} yields
\begin{align*}
    \frac{c}{2a} &= \sin \left(\!\frac{\alpha + \beta}{2}\!\right) \sin\left(\!\frac{\alpha - \beta}{2}\!\right), \\
    \frac{r_1 + r_2}{2a} &= 1 - \cos\left(\!\frac{\alpha + \beta}{2}\!\right) \cos\left(\!\frac{\alpha - \beta}{2}\!\right), \\
    \sqrt{\frac{GM}{a^{3}}}(t_2-t_1) &= 2\pi Q + \alpha - \beta -2\cos\left(\!\frac{\alpha + \beta}{2}\!\right) \sin\left(\!\frac{\alpha - \beta}{2}\!\right).
\end{align*}
We can now apply the trigonometric identities \eqref{eq:TrigIDs} to rewrite the system in terms of cosines and sines of a single angle (as opposed to sums and differences), obtaining
\begin{subequations}\label{eq:systemalphabeta}
\begin{align}
\label{eq:systemalphabetaA}
    \frac{c}{a} &= \cos\beta - \cos\alpha,\\
    \label{eq:systemalphabetaB}
    2 - \frac{r_1 + r_2}{a} &=\cos\beta +\cos\alpha,\\
    \label{eq:systemalphabetaC}
    \sqrt{\frac{GM}{a^{3}}}(t_2-t_1) &= 2\pi Q + \alpha - \sin\alpha -(\beta - \sin\beta).
\end{align}
\end{subequations}
The last equation of the system above is sometimes called \textit{Lagrange's transfer time equation}. Using the first two equations to solve for $\cos\alpha$ and $\cos\beta$ we obtain
\begin{equation}\label{eq:cosalpha}
\cos\alpha = 1 - \frac{r_1+r_2+c}{2a}  \qquad \text{ and } \qquad \cos\beta = 1 -\frac{r_1+r_2-c}{2a}\,.
\end{equation}
We will need to perform some further manipulations on the second of these equalities as follows
\[
\cos\beta = 1 -\frac{r_1+r_2-c}{2a} = 1 -\frac{(r_1+r_2)^2-c^2}{2a(r_1+r_2+c)}.
\]
Using the law of cosines, we can express the square of the length of the chord in terms of the true anomalies $\theta_1$ and $\theta_2$ as
\[
c^2 = r_1^2 -2r_1r_2\cos(\theta_2-\theta_1) + r_2^2\,,
\]
and substitute above to obtain
\begin{equation}\label{eq:cosbeta}
\cos\beta = 1 -\frac{r_1r_2(1+\cos(\theta_2-\theta_1))}{a(r_1+r_2+c)} = 1 -\frac{2r_1r_2\cos^2\left(\tfrac{1}{2}(\theta_2-\theta_1)\right)}{a(r_1+r_2+c)},
\end{equation}
where we made use of the identity $1+\cos A = 2\cos^2(A/2)$.

To connect the equality on the left of \eqref{eq:cosalpha} and \eqref{eq:cosbeta}(that involve only cosines) to equation \eqref{eq:systemalphabetaC} (that involves only sines), we will make use of the trigonometric identity ${\displaystyle \cos\theta = 1-2\sin^2(\theta/2)}$ which, upon substitution into said equations, yields
\begin{equation}\label{eq:sin^2a}
\sin^2(\alpha/2) = \frac{r_1+r_2+c}{4a} \qquad \text{ and } \qquad  \sin^2(\beta/2) = \frac{r_1r_2\cos^2\left(\tfrac{1}{2}(\theta_2-\theta_1)\right)}{a(r_1+r_2+c)} .
\end{equation} 
Solving for $1/a$ in the two expressions above, equating and taking square roots leads to
\begin{equation}\label{eq:sinasinb}
\sin(\beta/2) = \left(\frac{2(r_1r_2)^{1/2}\cos\left(\tfrac{1}{2}(\theta_2-\theta_1)\right)}{r_1+r_2+c}\right)\sin(\alpha/2),
\end{equation}
where the positive and negative roots are determined by the difference $\theta_2-\theta_1$, as the cosine will yield positive and negative values depending on whether $\theta_2-\theta_1$ belongs to the interval $[0,\pi]$ or $[\pi,2\pi]$.

We now apply the first equation in \eqref{eq:sin^2a} to $a$ to express the left hand side of equation \eqref{eq:systemalphabetaC} as
\[
\sqrt{\frac{GM}{a^{3}}}(t_2-t_1) = (t_2-t_1)\sqrt{GM}\left(\frac{2\sin(\alpha/2)}{\sqrt{r_1+r_2+c}}\right)^3.
\]
From this it follows that we can rewrite \eqref{eq:systemalphabetaC} as
\begin{equation}\label{eq:onlyalphabeta}
(t_2-t_1)\sqrt{GM}\left(\frac{2\sin(\alpha/2)}{\sqrt{r_1+r_2+c}}\right)^3 = 2\pi Q + \alpha - \sin\alpha -(\beta - \sin\beta).
\end{equation}
The nonlinear system defined by equations \eqref{eq:sinasinb} and \eqref{eq:onlyalphabeta} involves only the unknown auxiliary angles $\alpha$ and $\beta$. Since $r_1,r_2,\theta_1,\theta_2$, and $c$ are all known, it suffices to prescribe the travel time $t_2-t_1$ to be able to attempt a solution. Note that, depending on the values of the parameters given, no solutions may exist or there may be multiple solutions. In the particular case when less than one full orbit is completed during the travel time (i.e. $Q=0$), any existing solution must be unique \cite{Battin,simo1973,Woollands2017}. Once the system is solved for these two unknowns, the length of the semi--major axis $a$ can be recovered from the left equality in \eqref{eq:sin^2a}. This value, together with equation \eqref{eq:aenergy} produces the energy required to place the vessel on the elliptic transfer orbit. 

\subsection{A second historical digression and some closing remarks}

The polymath Johann Heinrich Lambert was born on August 26$^{th.}$ 1728 in M\"ulhausen, Alsace---which was then part of the Swiss confederation and now is the city of Mulhouse, France---to a modest family. He left formal schooling at the age of 12 to assist in his father's tailor business, but managed to find time to continue his studies independently. In 1764 he was appointed a Royal Professor in the Academy of Berlin, by the Emperor Friederich II of Prussia. He died on September 25$^{th.}$ 1777 in Berlin from respiratory complications---pneumonia according to some sources \cite{Volk1980} or tuberculosis according to others \cite{DorregoLpez2023}. A brief account of his life can be found at \cite{MacTutor}, while a more detailed one can be found in \cite{DorregoLpez2023}. Although with the passage of time his mathematical work was overshadowed by that of his illustrious contemporaries Leonhard Euler, Joseph--Louis Lagrange, and Pierre--Simon Laplace, during his lifetime he was counted among Europe's top philosophers, astronomers, physicists and mathematicians and became known as \textit{the Alsatian Newton} or \textit{the Alsatian Leibniz} \cite{Volk1980}. He made numerous contributions to number theory, geometry, and probability, and is credited with producing the first proof of the irrationality of $\pi$ \cite{Wallisser}. 

His interest in celestial mechanics dates back at least to 1744, when he got captivated by the great comet of Klinkenberg--Ch\'eseaux \cite{DorregoLpez2023} which---by contemporary accounts---developed as many as six tails. It is unclear if the young Lambert obtained any mathematical results in the immediate aftermath of the comet's passing, but we know that by early 1761 the idea behind Lambert's theorem---under the assumption of a parabolic trajectory---had matured in his mind enough for him to share it with Leonhard Euler in a letter dated on February 6$^{th.}$ 1761:
\begin{adjustwidth}{3cm}{0cm}
\textit{
``I forgot to turn problem §210 around, that the orbit may be found
$1^\circ$ by the 3 sides $FN$, $FM$, $NM$ and the time $T$ required to traverse the arc $NM$. $2^\circ$ by the ratio $(FM:FN)$, the angle $NFM$, the time $T$ and the periodic time. If the diameter of the Sun can be measured precisely enough, two observations suffice to determine the Earth’s orbit using the latter theorem."  \cite{Albouy2019,Bopp:1924,EulerArchive}
}
\end{adjustwidth}
Very shortly after his letter to Euler, Lambert went on to treat the elliptic and hyperbolic cases and published the results as part of his book \cite{Lambert:1761}. He sent the book to Euler who, in March 24$^{th.}$ 1761, replied in a letter:
\begin{adjustwidth}{3cm}{0cm}
\textit{
``The beautiful proof of the area of a parabolic sector, the expression for which you communicated to me gave me great pleasure; but I was even more surprised to see its application to elliptic sectors [...] I easily recognize that the methods I proposed earlier may be improved considerably."  \cite{Albouy2019,Bopp:1924,EulerArchive}
}
\end{adjustwidth}
As indicated by the final phrase above, in fact Euler had started working on the determination of the movement of comets at least as far back as 1742, after the passing of Halley's comet, publishing in 1743 a geometric method \cite{Euler1743} and then in 1744 the first purely analytic method to determine a parabolic orbit based on three observations of the comet \cite{Bistafa2021,Euler1744}. However, Euler did not attempt to generalize his calculations to orbits other than parabolic and therefore Lambert is generally credited with the introduction of the problem that now bears his name \cite{Albouy2019}.

In his proof, Lambert made use solely of geometric tools. This made the argument hard to understand and even harder to apply for practical calculations. In fact, in the same letter mentioned above, Euler points out that:
\begin{adjustwidth}{3cm}{0cm}
\textit{
``Your theorem for expressing the area of a parabolic sector is excellent, I can see the truth of it, but by such
detours, that I could never have arrived to it had I not known it in advance; I therefore wait impatiently to see
the analysis leading to it without detours."  \cite{Albouy2019,Bopp:1924,EulerArchive}
}
\end{adjustwidth}
Lambert did not arrive at an alternate proof using the modern methods of calculus. This would have to wait until 1780, already after his passing, when Lagrange finally arrived at an argument similar to the one presented here. Countless other proofs to the theorem have been given and rediscovered over time. In the wonderful---but heavily mathematical---article \cite{Albouy2019}, Albouy discusses several ``families" of proofs and presents a detailed timeline of early attempts and some highlights of modern ones. 

In his original statement, Lambert in fact did not pose a problem in the way we did at the beginning of Section \ref{sec:Lambert}. Instead, he made a statement about the relevant variables that determine the travel time, claiming that it depends only on the distances to the points, the length of the chord connecting them and the length of the semi--major axis of the conic (the parabolic case can be obtained by letting $a\to\infty$ \cite[Problem 4.3]{CoPr1993}). In modern notation, the theorem can be stated as:
\begin{theorem}[Lambert's theorem]\label{thm:Lambert}
Let $P_1$ and $P_2$ denote the positions of a particle moving under the influence of a gravitational potential at times $t_1$ and $t_2$. If $r_1,r_2$ denote the distances of $P_1$ and $P_2$ from the attractive center and $c:=|P_2-P_1|$, there exists a, possibly multi--valued, function such that
\[
t_2-t_1 = F(a,r_1+r_2,c).
\]
\end{theorem}
Going back to the analysis in the previous section, and recalling that the auxiliary variables $\alpha$ and $\beta$ are both functions of $r_1+r_2$, $c$ and the semi--major length $a$, it is clear that equation \eqref{eq:onlyalphabeta} is only one step away from the form appearing in Theorem \ref{thm:Lambert}, which is why Lagrange's argument constitutes an analytic proof of Lambert's original statement. 

Although clearly Theorem \ref{thm:Lambert} is related to the orbit--determination Problem \ref{prob:Lambert}, they are certainly not the same thing. According to Albouy \cite{Albouy2019}, it was not until the 1960's when the terms \textit{Lambert's theorem} and \textit{Lambert's problem} started being used interchangeably---with the second one eventually taking over. This was perhaps due to the fact that, with the advent of the space age, Lambert's work was seen through the lens of orbital determination algorithms, many of which involved the solution of Lambert's problem \ref{prob:Lambert}. As it concerns practical solution methods---rather than proofs of the theorem---there have been myriads of proposed algorithms. A recent comparative study of several of the most widely used approaches can be found in \cite{MG2021}. 
%
\section{Statements and declarations}
%

\textbf{Author contribution statment.}  Because of the nature of mathematical research, authors are considered on equal footing and are always listed in alphabetical order. In other words, \emph{as opposed to other disciplines, in mathematics there is no ``first author"}. Lenox Helene Baloglou, Parneet Gill and Tonatiuh S\'anchez-Vizuet all contributed in equal proportion.

\textbf{Competing interests.} The authors declare that they have no conflict of interest.

\textbf{Funding.} All authors were partially funded by the United States National Science Foundation through the grant NSF-DMS-2137305. Lenox Helene Baloglou is thankful for the generous support of the University of Arizona's RII-Sponsored Campuswide Undergraduate Student--Initiated Original Research Program.

%
\bibliographystyle{abbrv}
\bibliography{references}

\end{document}